\documentclass[prd,twocolumn,superscriptaddress,altaffilletter,amssymb,showpacs,nofootinbib]{revtex4}
\usepackage[dvips]{graphicx}
\usepackage{amsmath}
\newcommand{\be}{\begin{equation}}
\newcommand{\ee}{\end{equation}}
\newcommand{\bea}{\begin{eqnarray}}
\newcommand{\eea}{\end{eqnarray}}

\begin{document}

\title{Self-interacting Scalar Field Trapped in a DGP Brane: The Dynamical Systems Perspective}

\author{Israel Quiros}\email{israel@uclv.edu.cu}
\affiliation{Departamento de F\'{\i}sica, Universidad Central de
Las Villas, 54830 Santa Clara, Cuba.}\affiliation{Part of the Instituto
Avanzado de Cosmolog\'ia (IAC) collaboration http://www.iac.edu.mx/}
\author{Ricardo Garc\'{\i}a-Salcedo}\email{rigarcias@ipn.mx}
\affiliation{Centro de Investigacion en Ciencia Aplicada y Tecnologia Avanzada - Legaria del IPN, M\'exico D.F., M\'exico.} \author{Tonatiuh Matos}\email{tmatos@fis.cinvestav.mx}
\affiliation{Departamento de F{\'\i}sica, Centro de Investigaci\'on y de Estudios Avanzados del IPN, A.P. 14-740, 07000 M\'exico D.F., M\'exico.} \affiliation{Part of the Instituto
Avanzado de Cosmolog\'ia (IAC) collaboration http://www.iac.edu.mx/}
\author{Claudia Moreno}\email{claudia.moreno@cucei.udg.mx}
\affiliation{Departamento de F\'{\i}sica y Matem\'aticas, Centro Universitario de Ciencias Ex\'actas e Ingenier\'{\i}as, Corregidora 500 S.R., Universidad de Guadalajara, 44420
Guadalajara, Jalisco, M\'exico} \affiliation{Part of the Instituto Avanzado de Cosmolog\'ia (IAC) collaboration http://www.iac.edu.mx/}

\date{\today}
\begin{abstract}
We apply the dynamical systems tools to study the linear dynamics of a self-interacting scalar field trapped on a DGP brane. The simplest kinds of self-interaction potentials are investigated: a) constant potential, and b) exponential potential. It is shown that the dynamics of DGP models can be very rich and complex. One of the most interesting results of this study shows that dynamical screening of the scalar field self-interaction potential, occuring within the Minkowski cosmological phase of the DGP model and mimetizing 4D phantom behaviour, is an attractor solution for a constant self-interaction potential but not for the exponential one. In the latter case gravitational screening is not even a critical point of the corresponding autonomous system of ordinary differential equations.
\end{abstract}

\pacs{04.20.-q, 04.20.Cv, 04.20.Jb, 04.50.Kd, 11.25.-w, 11.25.Wx,
 95.36.+x, 98.80.-k, 98.80.Bp, 98.80.Cq, 98.80.Jk}
\maketitle

\section{Introduction}

Since the discovery that our universe can be currently undergoing a stage of accelerated expansion \cite{obs}, many phenomenological models based either on Einstein General Relativity (EGR), or using alternatives like the higher dimensional brane world theories \cite{roy}, have been invoked (for a recent review on the subject see reference \cite{Copeland:2006wr}). The latter ones, being phenomenological in nature, are inspired by string theory.

One of the brane models that have received most attention in recent years is the so called Dvali-Gabadadze-Porrati (DGP) brane world \cite{dgp}.\footnotemark\footnotetext{For cosmology of DGP braneworlds see reference \cite{dgpcosmology}.} This model describes a brane with 4D world-volume, that is embedded into a flat 5D bulk, and allows for infrared (IR)/large scale modifications of gravitational laws. A distinctive ingredient of the model is the induced Einstein-Hilbert action on the brane, that is responsible for the recovery of 4D Einstein gravity at moderate scales, even if the mechanism of this recovery is rather non-trivial \cite{deffayet}. The acceleration of the expansion at late times is explained here as a consequence of the leakage of gravity into the bulk at large (cosmological) scales, so it is just a 5D geometrical effect,unrelated to any kind of misterious "dark energy". 

As with many IR modifications of gravity, there are ghosts modes in the spectrum of the theory \cite{kazuya,koyama}.\footnotemark\footnotetext{In fact there are ghosts only in one of the branches of the DGP model; the so called "self-accelerating" branch, or self-accelerating cosmological phase \cite{lue}.} Nevertheless, studying the dynamics of DGP models continues being a very atractive subject of research. It is due, in part, to the very simple geometrical explanation to the "dark energy problem", and, in part, to the fact that it is one of a very few possible consistent IR modifications of gravity that might be ever found. 

The aim of this letter is, precisely, to study the dynamics of a self-interacting scalar field trapped on a DGP brane \footnotemark\footnotetext{Regarding the so called "normal" branch, or Minkowski cosmological phase \cite{lue} of DGP model, this is just a particular member in the wider class of models of reference \cite{clmq}.} by invoking the dynamical systems tools, which have been proved useful to retrieve significant information about the evolution of a huge class of cosmological models (see for instance the book \cite{coley}).  The simplest self-interaction potentials: a) constant potential, and b) exponential potential, are investigated. The constant potential is the simplest one, and, for frozen scalar field $\phi=\phi_0$, for the Minkowski cosmological phase, the corresponding model coincides with the one of reference \cite{lue}. The exponential potential represents a common functional form for self-interaction potentials that can be found in higher-order \cite{witt} or higher-dimensional theories \cite{nd}. These can also arise due to non-perturbative effects \cite{nonp}. In addition to the scalar field we consider, also, a background fluid trapped on the DGP brane.

A dynamical study of DGP models with a scalar field trapped on the DGP brane has been already undertaken in reference \cite{zhang} for an exponential potential, to show that crossing of the phantom barrier is indeed possible in DGP cosmology with a single scalar field (see also \cite{clmq} in this regard). However, the authors of that paper do not study in detail the phase space of the model and, in correspondence, they are not able to find critical points. Their claim that scaling solutions do not exist in a universe with dust on a DGP brane, seems to be in contradiction with our results. Also, in reference \cite{gumjudpai} a dynamical systems study of DGP scenarios is undertaken for several self-interaction potentials. However the author of the latter reference considers only a scalar field trapped on the brane (no other kind of matter is trapped on it). Besides, phase space variables used in \cite{gumjudpai} are different from the ones used in the present study.

In a sense, the present investigation is a continuation of the one reported in paper \cite{wands}, to include higher-dimensional behaviour (in the present case dictated by the DGP dynamics). In consequence, for the exponential self-interaction potential, our results will include the ones reported in \cite{wands} as a particular case.

Through the paper we use natural units ($8\pi G=8\pi/m_{Pl}^2=\hbar=c=1$).

\section{The Model}

As already mentioned, we will be concerned here with a DGP brane model where self-interacting scalar field matter is trapped on the DGP brane. The field equations are the following:
\bea
Q_{\pm}^2=\frac{1}{3}(\rho+\frac{1}{2}\dot\phi^2+V(\phi)),\nonumber\\
\dot\rho=-3\gamma H\rho,\;\;\ddot\phi=-3H\dot\phi-\partial_\phi
V,\label{feqs}
\eea
where $\rho$ is the energy density of the background barotropic fluid ($\gamma$ is the barotropic index), $\phi$ is the scalar field trapped in the DGP brane, $V$ its self-interaction potential, and we have used the following definition:
\be
Q_{\pm}^2\equiv H^2\pm\frac{1}{r_c}\;H,\label{qpm}
\ee
with $r_c$ being the so called crossover scale. Depending on the choice of the signs "+" or "-" in (\ref{qpm}), there are two possible branches of the DGP model, corresponding to two possible embeddings of the DGP brane in the Minkowski bulk. The choice "+" is for the so called Minkowski cosmological phase of DGP that is free of ghosts, while the "-" choice is for the self-accelerating cosmological phase. The Minkowski phase of the present model belongs in the wider class of models of reference \cite{clmq}. For $\phi=const$, the self-accelerating phase coincides with the model of \cite{lue}.

\section{The Phase Space}

Our aim is to write the system (\ref{feqs}) as an autonomous system of ordinary differential equations. We make the following choice of dynamical variables:\footnotemark\footnotetext{In what follows, for brevity, we avoid writing the "$\pm$" sign.}
\be
x=\frac{1}{\sqrt{6}}\frac{\dot\phi}{Q},\;\;y=\frac{1}{\sqrt{3}}\frac{\sqrt
 V}{Q},\;\;z=\frac{Q}{H},\label{dvars}
\ee
and, also, we introduce the time variable $\tau=\int H dt$. The following autonomous system of ordinary differential equations is obtained:
\bea &&x'=-3x-\sqrt\frac{3}{2}\;(\partial_\phi\ln
 V)\;y^2z+\nonumber\\
&&\;\;\;\;\;\;\;\;\;\;\;\;\;\;\;\;\;\;\;\;\;\;
\;\;\;\;\;\;\;\;\;\;\;\;\;\;\;\frac{3}{2}x(\gamma(1-x^2-y^2)+2x^2),\nonumber\\
&&y'=\sqrt\frac{3}{2}\;(\partial_\phi\ln
 V)\;xyz+\frac{3}{2}\;y(\gamma(1-x^2-y^2)+2x^2),\nonumber\\
&&z'=\frac{3}{2}\;z\left(\frac{z^2-1}{z^2+1}\right)(\gamma(1-x^2-y^2)+2x^2),\label{asode}
\eea where a prime denotes derivative with respect to $\tau$. In terms of the dynamical variable $z$, equation (\ref{qpm}) can be rewritten as:
\be z^2=1\pm\frac{1}{r_c H}.
\ee
For the Minkowski phase, since $0\leq H\leq\infty$ (we consider just non-contracting universes), then $1\leq z\leq\infty$. The case $-\infty\leq z\leq -1$ corresponds to the time reversal of the latter situation. For the self-accelerating phase, $-\infty\leq z^2\leq 1$, but since we want real valued $z$ only, then $0\leq z^2\leq 1$.\footnotemark\footnotetext{In fact, fitting SN observations requires $H\geq r_c^{-1}$ in order to achieve late time acceleration (see, for instance, reference \cite{koyama} and references therein). This means that $z$ has to be real-valued.} As before, the case $-1\leq z\leq 0$ represents time reversal of the case $0\leq z\leq 1$ that will be investigated here. Both branches share the common subset $(x,y,z=1)$, which corresponds to the formal limit $r_c\rightarrow\infty$ (see equation (\ref{qpm})), i.e., this represents just the standard 4D behaviour of (4D) Einstein-Hilbert theory oupled to a self-interacting scalar field.

In terms of the above introduced dynamical variables $x,y,z$, the "effective" dimensionless density parameters $\bar\Omega=\rho/3Q^2$, and $\bar\Omega_\phi=\rho_\phi/3Q^2$ ($\rho_\phi=\dot\phi^2/2+V$) can be expressed in the following way:

\be
\bar\Omega=\frac{\Omega}{z^2}=1-x^2-y^2,\;\;\bar\Omega_\phi=\frac{\Omega_\phi}{z^2}=x^2+y^2,
\ee where, as customary, $\Omega=\rho/3H^2$, $\Omega_\phi=\rho_\phi/3H^2$. Since both $\Omega\geq 0$, $\Omega_\phi\geq 0$, and $z^2\geq 0$, then $0\leq x^2+y^2\leq 1$.
This means the phase space for the dynamical system driven by the evolution equations (\ref{asode}), for the Minkowski cosmological phase (the ghost-free "+" branch), is given by the unbounded region:

\be
\Psi_+=\{(x,y,z):\;0\leq x^2+y^2\leq 1,\;z\in [1,\infty[\},
\ee while, for the self-accelerating cosmological phase (the pathological "-" branch), it is given by the complementary (non-compact) region:

\be
\Psi_-=\{(x,y,z):\;0\leq x^2+y^2\leq 1,\;z\in ]0,1]\}.
\ee

Notice that the points belonging in the set $(x,y,0)$ can not be included since, in this case ($z=0\;\Rightarrow\;Q=0$) the variables $x$ and $y$ are undefined. In consequence, the self-accelerating solution $H=1/r_c$ has to be estudied separately, i. e., with a different choice of phase space variables (see section \ref{selfa}).

Take a look at (\ref{asode}). These equations hide the differences  between the "+" and "-" branches of the DGP. Nevertheless, the differences indeed exist. These are encoded in the different definition of the corresponding phase spaces (see the definitions for $\Psi_+$ and $\Psi_-$ above) which, from now on, we will call as Minkowski ($\Psi_+$) and self-accelerating ($\Psi_-$) phases respectivelly. Recall that points in the phase plane $\{(x,y,1):0\leq x^2+y^2\leq 1\}$, where both $\Psi_+$ and $\Psi_-$ intersect, are related to standard 4D behaviour (formal limit $r_c\rightarrow\infty$). 

To finalize this section we write another useful magnitudes of observational interest in terms of the dynamical variables (\ref{dvars}). These are, the deceleration parameter $q=-(1+\dot H/H^2)$:

\be
q=-1+\frac{3z^2}{z^2+1}(\gamma(1-x^2-y^2)+2x^2),\label{q}
\ee and the equation of state (EOS) parameter $\omega_\phi=(\dot\phi^2-2V)/(\dot\phi+2V)$:

\be
\omega_\phi=\frac{x^2-y^2}{x^2+y^2}.\label{eos}
\ee

\begin{table*}[tbp]\caption[crit]{Properties of the critical points for the autonomous system (\ref{asode1}).}
\begin{tabular}{@{\hspace{4pt}}c@{\hspace{14pt}}c@{\hspace{14pt}}c@{\hspace{14pt}}c@{\hspace{14pt}}c@{\hspace{14pt}}c@{\hspace{14pt}}c@{\hspace{14pt}}c}
\hline\hline\\[-0.3cm]
$P_i$ &$x$&$y$&$z$&Existence& $\bar\Omega_\phi$& $\omega_\phi$& $q$\\[0.1cm]
\hline\\[-0.2cm]
$P_1$& $0$&$0$&$1$ & Always ($\forall\gamma\in[0,2]$) & $0$&undefined&
$-1+\frac{3\gamma}{2}$\\[0.2cm]
$P_2^\pm$& $\pm 1$&$0$&$1$ & " & $1$&$1$& $2$\\[0.2cm]
$P_3^\pm$& $0$&$\pm 1$&$1$ & " & $1$&$-1$& $-1$\\[0.2cm]
$M_\pm$& $0$&$\pm 1$&$z\in ]0,\infty[$ & " & $1$&$-1$&$-1$\\[0.4cm]

\hline \hline
\end{tabular}\label{tab1}
\end{table*}
\begin{table*}[tbp]\caption[eigenv]{Eigenvalues for the critical points in table \ref{tab1}.} \begin{tabular}{@{\hspace{4pt}}c@{\hspace{14pt}}c@{\hspace{14pt}}c@{\hspace{14pt}}c@{\hspace{14pt}}c@{\hspace{14pt}}c@{\hspace{14pt}}c}
\hline\hline\\[-0.3cm]
$P_i$ &$x$&$y$& $z$& $\lambda_1$& $\lambda_2$& $\lambda_3$\\[0.1cm]\hline\\[-0.2cm]
$P_1$& $0$&$0$& $1$& $-3(2-\gamma)/2$& $3\gamma/2$&$3\gamma/2$\\[0.2cm]
$P_2^\pm$& $\pm 1$&$0$& $1$& $3(2-\gamma)$& $3$& $3$\\[0.2cm]
$P_3^\pm$& $0$&$\pm 1$& $1$& $-3$& $-3\gamma$& $0$\\[0.2cm]
$M_\pm$& $0$&$\pm 1$& $z$& $-3$& $-3\gamma$& $0$\\[0.4cm]
\hline \hline
\end{tabular}\label{tab2}
\end{table*}

\section{The Critical Points}

We study here two concrete yet generic cases. First we consider a constant potential $V=V_0$, and then we study the exponential potential $V=V_0\exp{(-\alpha\phi)}$. For more general classes of potentials we have to rely on a quite different approach (see, for instance reference \cite{llq}). This will be the subject of future research.

\subsection{The Constant Potential $V=V_0$}\label{v0}

The autonomous system of ordinary differential equations (\ref{asode}) simplifies to:
\bea
&&x'=-3x+\frac{3}{2}x(\gamma(1-x^2-y^2)+2x^2),\nonumber\\
&&y'=\frac{3}{2}y(\gamma(1-x^2-y^2)+2x^2),\nonumber\\
&&z'=\frac{3}{2}z(\frac{z^2-1}{z^2+1})(\gamma(1-x^2-y^2)+2x^2).\label{asode1}
\eea

Before finding the critical points of this system of equations and investigating their stability, a straightforward inspection of the diferential equations (\ref{asode1}) reveals that the first two equations (differential equations for $x$ and $y$ variables respectively) are decoupled from the third one. This means that there can be found critical points with $x=0$, $y=\pm 1$ and arbitrary $z$-s to form critical submanifolds of particular interest (see below on the critical submanifolds $M_\pm$). These critical points are associated with the inflationary ($q=-1$) DGP-Friedmann equation (notice that, in this case, $\Omega=0$, $\Omega_\phi=z^2$):

\be H^2\pm\frac{1}{r_c}\;H=\frac{V(\phi)}{3}.\label{dgpfriedmann}\ee For the Minkowski cosmological phase ("+" sign in (\ref{dgpfriedmann})), this solution can be associated with 4D phantom-like behaviour, that is generated through dinamical (gravitational) screening of the potential energy $V(\phi)$.\footnotemark\footnotetext{The present case ($V=V_0$) coincides with the case studied in \cite{lue}.} As we will see in the next sections, the latter solution is an attractor only for the constant potential $V=V_0$. For the exponential potential it is not even a critical point of the corresponding autonomous system of ordinary differential equations.

The critical points of the autonomous system of equation (\ref{asode1}) are summarized in table \ref{tab1}, while table \ref{tab2} shows the eigenvalues of the corresponding Jacobian matrices. Notice, from (\ref{asode1}), that the dynamical equations are invariant under the change of sign $z\rightarrow -z$, in consequence we have not included the points with $z=-1$ in our analysis.

There are 5 critical points $P_1$, $P_2^\pm$, $P_3^\pm$ and two critical submanifolds $M_\pm=(0,\pm 1,z)$ ($z\in ]0,1]$ for the self-accelerating phase, while, $z\in [1,\infty[$ for the Minkowski phase). It is obvious that critical points $P_3^\pm$ belong in $M_\pm$.

Notice that for critical points $P_1$, $P_2^\pm$, $P_3^\pm$, since $z=1$ (formal limit $r_c\rightarrow\infty$), then $\bar\Omega=\Omega$, $\bar\Omega_\phi=\Omega_\phi$, i.e, these points correspond to standard 4D behaviour. Otherwise, critical points attached to the phase plane $\{(x,y,1): 0\leq x^2+y^2\leq 1\}$ are the well known critical points for, basically, standard 4D Einstein-Hilbert gravity with a minimally coupled self-interacting scalar field with constant self-interaction potential. These points exist for both branches of the DGP model.

The matter-dominated solution $P_1$ ($\bar\Omega=\Omega=1$) is a saddle point in $\Psi_\pm$. It corresponds to an accelerating solution for $\gamma<2/3$ (decelerating otherwise, i.e, for $2/3\leq\gamma\leq 2$). Stiff-matter dominated solutions $P_2^\pm$ ($\bar\Omega_\phi=\Omega_\phi=1$, $\omega_\phi=1$) represent source (repellor) critical points in phase space, and are always decelerating ($q=2$). 

The scalar field dominated inflationary phases ($\bar\Omega_\phi=\Omega_\phi=1$, $\omega_\phi=-1$, $q=-1$) corresponding to points in the submanifolds $M_\pm$ in table \ref{tab1} (as already said, these include the critical points $P_3^\pm$), represent non-hyperbolic critical points, since one of the eigenvalues of the corresponding Jacobian matrixes vanishes (see table \ref{tab2}). In this case the only thing we can state with certainty, on the basis of straightforward analisys of the autonomous system of equations (\ref{asode1}), is that trajectories in phase space, originating in one of the repellor points $P_2^\pm$, depending on the phase considered: the Minkowski phase or the self-accelerating one, and on the initial conditions, will inevitably approach one or several of the points $(0,\pm 1, z_{i0})\in M_\pm$. Otherwise, points in the segments $(0,\pm 1,z)$, that are parallel to the $z$-axis, can be seen as attractors by the corresponding phase space trajectories. Actually, the stable subspace of the dynamical system (\ref{asode1}), near of the critical points in the submanifolds $(0,\pm 1,z)$, is spanned by the eigenvectors:

\bea
v_1=\left(\begin{array}{clrrr}
1\\
0\\0 \end{array}\right),\;\;v_2=\left(\begin{array}{clrrr}
0\\
1\\0 \end{array}\right),\nonumber
\eea i.e., the stable subspace is the plane $(x,y)$ intersecting the $z$-axis at a given $z=z_{i0}$. The central subspaces coincide with the open segments ($0,\pm 1,z\in ]0,1]$) for the self-accelerating phase, while, for the Minkowski phase these coincide with the semi-infinite segments ($0,\pm 1,z\in [1,\infty[$). These segments are parallel to the $z$-axis and lie on the boundary of $\Psi_\pm$ ($\partial\Psi_\pm$). Since, in this case, there is no unstable subspace related to $M_\pm$, and the centre manifold belongs in the boundary of the phase space, this means that trajectories in phase space that evolve from the stiff-matter repellor $P_2^\pm\in (x,y,1)$, will inevitably approach one or several points (depending on the cosmological phase being considered) in $M_\pm$, into the future ($\tau\rightarrow\infty$). As we will see quite soon, it is a fact that, only for the Minkowski phase $\Psi_+$, the points in $M_\pm$ are attractor critical points, so that, the corresponding submanifolds are attractor segments. These critical subspaces are an important result, since, as we will see in the next subsection, they exist only for the constant potential case, meaning that the associated cosmological solutions are generic only in this particular case.

\begin{figure}[ht]
\begin{center}
\includegraphics[width=5.5cm,height=5cm]{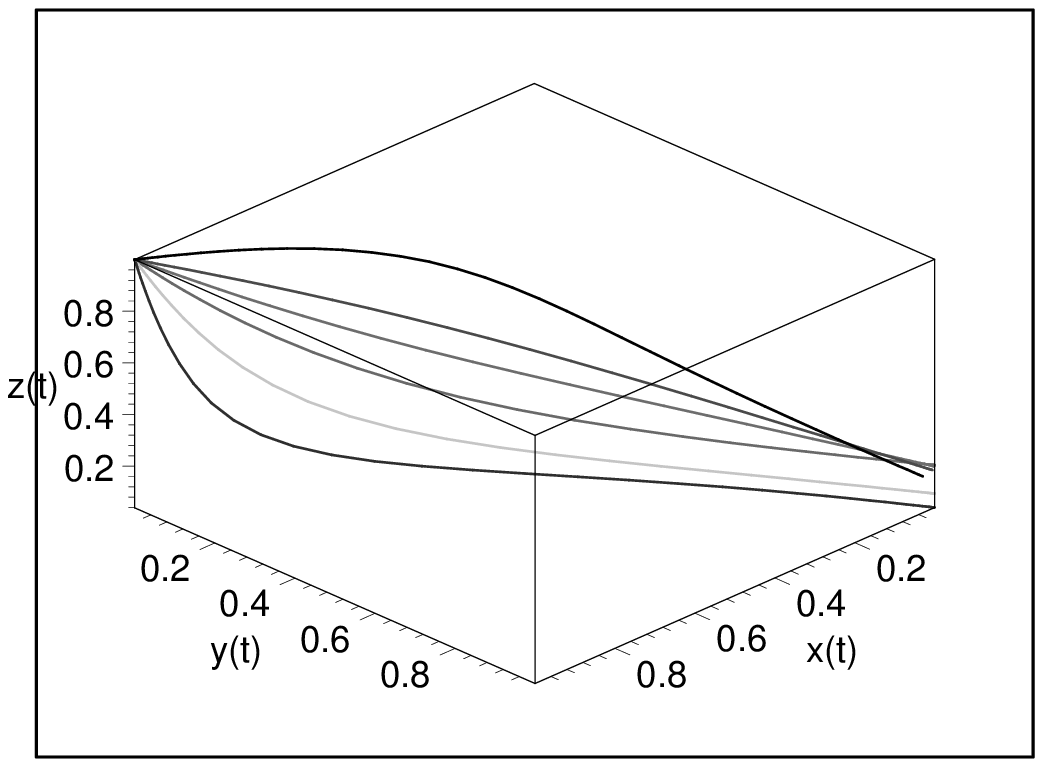}
\includegraphics[width=5.5cm,height=5cm]{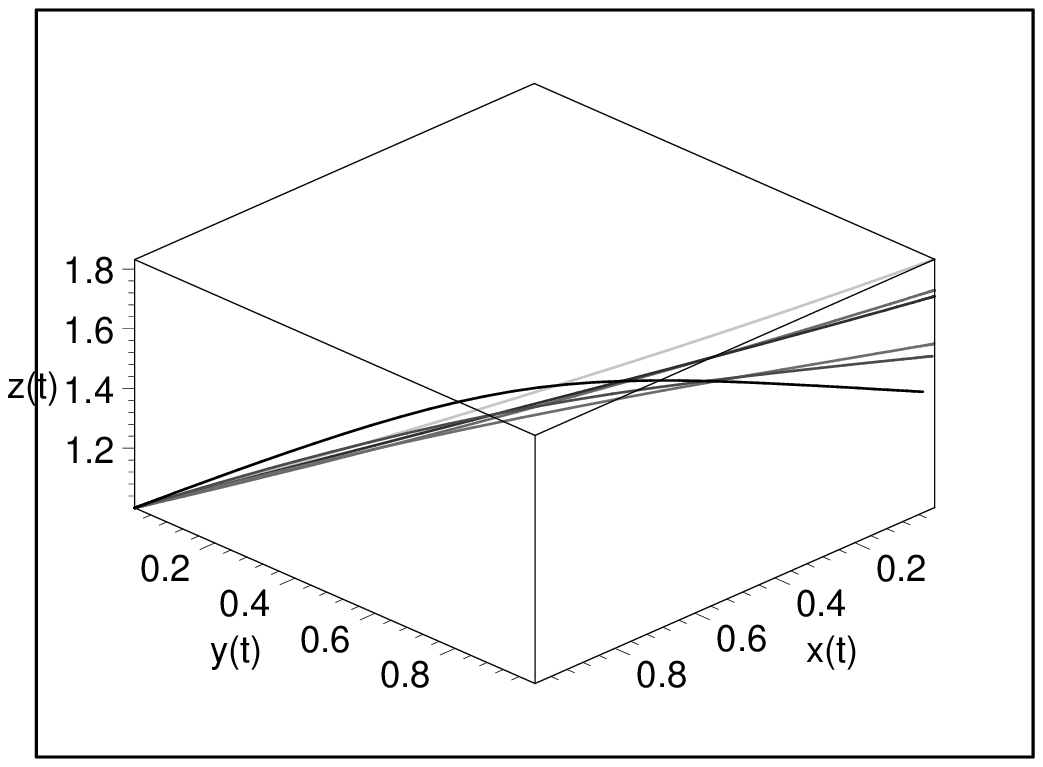}
\includegraphics[width=5.5cm,height=5cm]{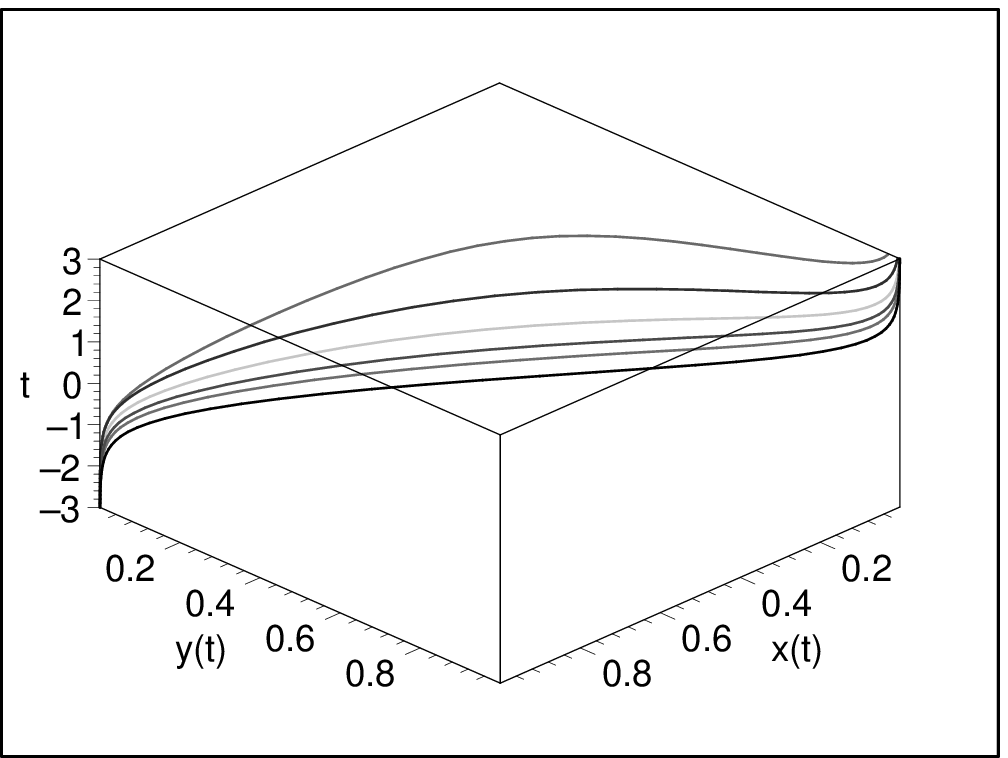}
\vspace{0.3cm}
\caption{Trajectories in phase space for different sets of initial conditions for the constant potential $V=V_0$. The upper figure and the figure at the center are for the self-accelerating $\Psi_-$ and Minkowski $\Psi_+$ phases of the DGP model respectively. The figure at the botton shows the flow in time $\tau$. Notice that trajectories in $\Psi_-$ asymptotically approach to the self-accelerating solution $H=1/r_c$. Meanwhile, the trajectories in $\Psi_+$ approach points in the segment $(0,1,z)$.}\label{fig01}
\end{center}
\end{figure}

Additional and, perhaps, more precise information can be extracted from the phase portraits. Actually, in the figure \ref{fig01}, phase trajectories are shown for the constant potential $V=V_0$, for different sets of initial conditions. The upper figure is for the self-accelerating phase $\Psi_-$, while the figure in the middle is for the Minkowski phase $\Psi_+$. The lower figure shows the flow of the autonomous system (\ref{asode1}). From these figures several features are apparent:

\begin{itemize}

\item Phase trajectories in $\Psi_-$ (upper figure in fig. \ref{fig01}) originate from the source critical points $P_2^\pm$ (only the point $P_2^+$ is shown), correponding to the standard 4D kinetic energy dominated (stiff-matter) solution, and (asymptotically) approach to the point $(0,1,0)$ that has been removed from the phase space since phase space variables $x$ and $y$ blow up at the phase plane ($x,y,0$). The dynamics in the neigbourhood of this point has to be investigated in terms of different phase space variables. Therefore, for the self-accelerating cosmological phase $\Psi_-$ the points $(0,\pm 1,z_{i})\in M_\pm$ are not even critical points of the autonomous system (\ref{asode1}).

\item Phase trajectories in $\Psi_+$ (center figure in fig. \ref{fig01}) originate from the 4D stiff-matter solution (unstable node $P_2^+$ in tab. \ref{tab1}) and end up at the inflationary points $(0,1,z_{0i})\in M_+$, where the different $z_{0i}$-s are associated with the different initial conditions. Otherwise, points in $M_+$ are seen as attractor points by the different phase space "observers", moving along different phase trajectories that originate at $P_2^+$. These scalar field dominated critical points correspond to inflationary solutions of the DGP-Friedmann equation: $$H^2+\frac{1}{r_c}\;H=\frac{V_0}{3}.$$ 

\end{itemize}

That the phase trajectories in $\Psi_+$ (middle figure in fig. \ref{fig01}), leave the phase plane $(x,y,1)$ and probe the bulk of the phase space $\Psi_+$, is a nice illustration of the fact that 4D phantom-like behaviour arising from dynamical screening of the brane cosmological constant $V_0$ \cite{lue}, is a phenomenom of 5D nature.\footnotemark\footnotetext{This argument has been suggested to us by Roy Maartens.} The new feature revealed by the present dynamical systems analysis relies on the fact that this type of behaviour -- gravitational screening -- is quite independent of the initial conditions, since the corresponding semi-infinite segments $M_\pm\in\partial\Psi_+$ are stable submanifolds.

\begin{table*}[tbp]\caption[crit1]{Critical points for the autonomous system (\ref{asode2}).}
\begin{tabular}{@{\hspace{4pt}}c@{\hspace{14pt}}c@{\hspace{14pt}}c@{\hspace{14pt}}c@{\hspace{14pt}}c@{\hspace{14pt}}c@{\hspace{14pt}}c@{\hspace{14pt}}c}
\hline\hline\\[-0.3cm]
$P_i$ &$x$&$y$&$z$&Existence& $\bar\Omega_\phi$&
 $\omega_\phi$& $q$\\[0.1cm]\hline\\[-0.2cm]
$P_1$& $0$&$0$&$1$ & All $\alpha$, $\gamma$ & $0$ &undefined&$-1+\frac{3\gamma}{2}$\\[0.2cm]
$P_{2,3}$& $\pm 1$&$0$&$1$ & All $\alpha$, $\gamma$ &$1$&$1$& $2$\\[0.2cm]
$P_4$& $\frac{\alpha}{\sqrt 6}$&$\sqrt{1-\frac{\alpha^2}{6}}$&$1$ & $\alpha^2<6$& $1$&$\frac{\alpha^2}{3}-1$& $\frac{\alpha^2}{2}-1$\\[0.2cm]
$P_{5}$&$\sqrt\frac{3}{2}\frac{\gamma}{\alpha}$&$\sqrt\frac{3(2-\gamma)\gamma}{2\alpha^2}$&$1$ & $\alpha^2>3\gamma$ &$\frac{3\gamma}{\alpha^2}$&$\gamma-1$& $-1+\frac{3\gamma}{2}$\\[0.4cm]\hline \hline
\end{tabular}\label{tab3}
\end{table*}

\begin{figure}[t]
\begin{center}
\includegraphics[width=5.5cm,height=5cm]{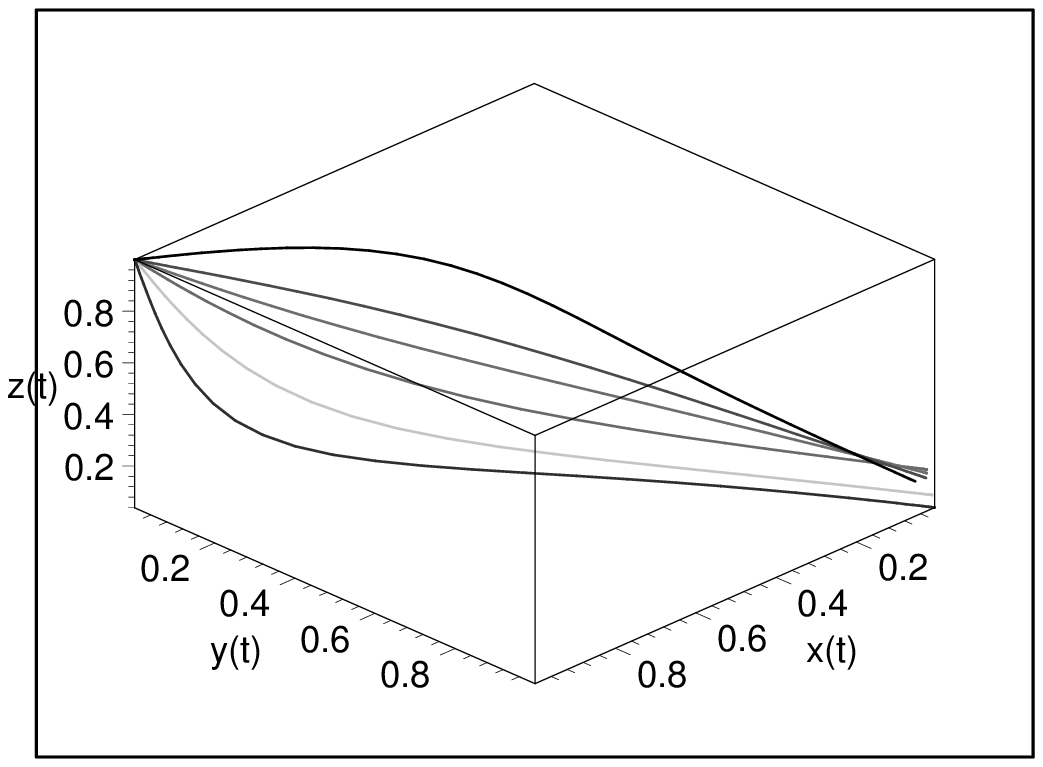}
\includegraphics[width=5.5cm,height=5cm]{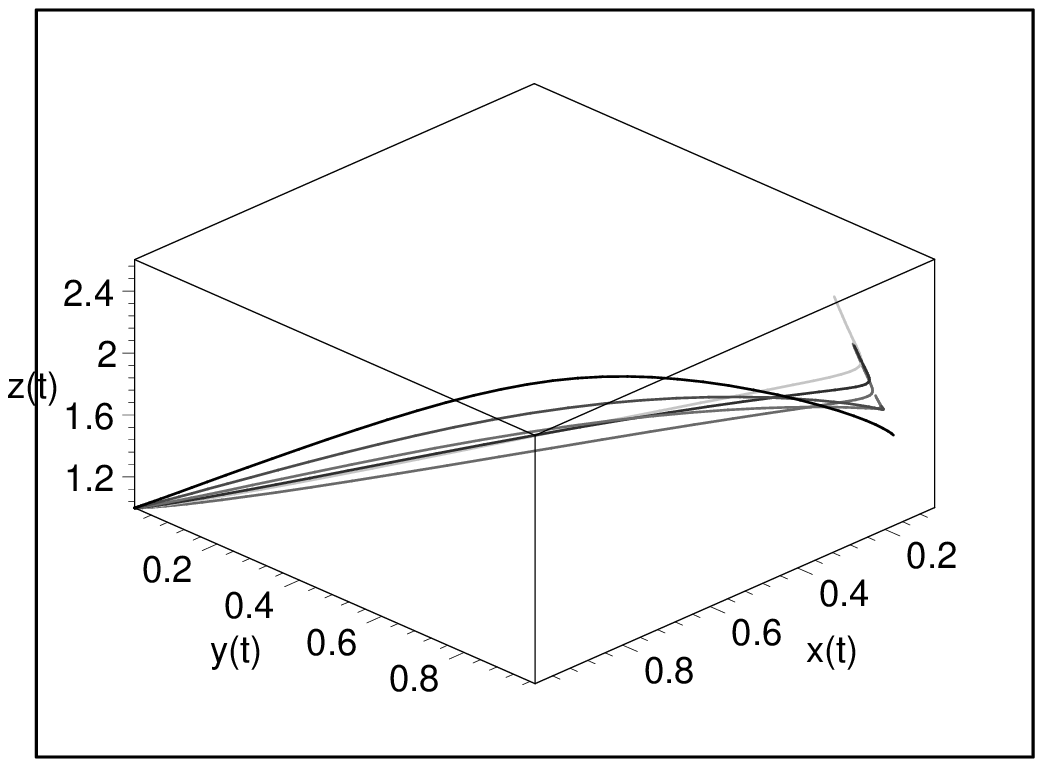}
\includegraphics[width=5.5cm,height=5cm]{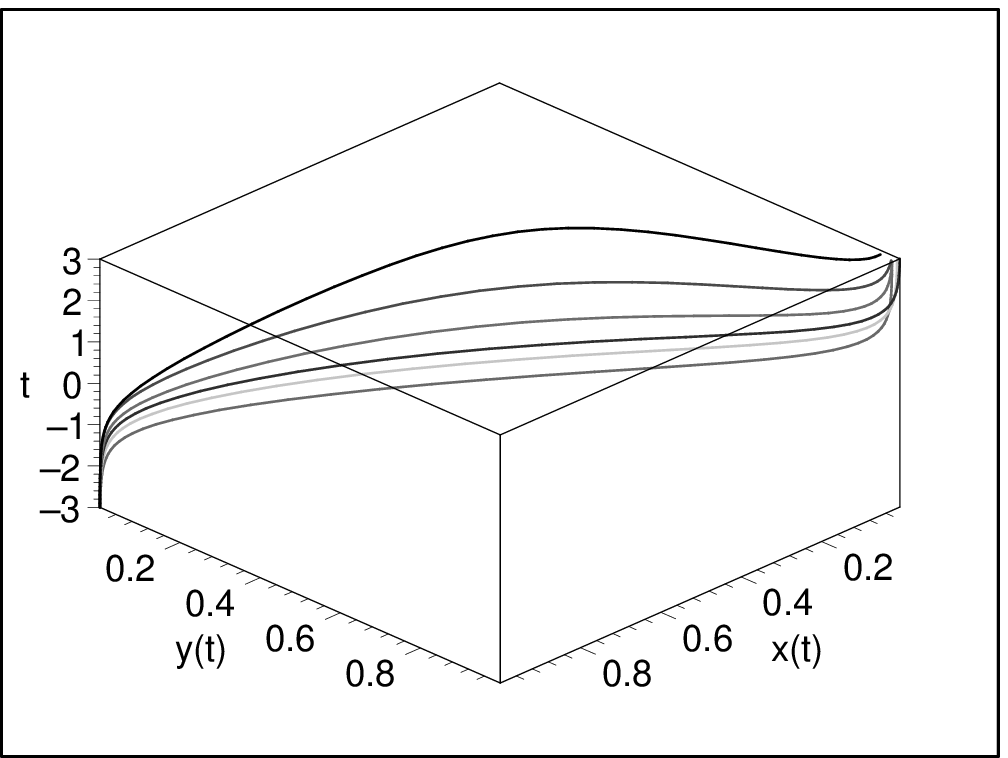}
\vspace{0.3cm}
\bigskip
\caption{Phase trajectories for different sets of initial conditions for the exponential potential $V=V_0 \exp(-\alpha\phi)$. As before the upper figure and the figure at the center are for the self-accelerating $\Psi_-$ and Minkowski $\Psi_+$ phases of the DGP model respectively. The flow of the autonomous system in time $\tau$ is shown in the figure at the botton. Unlike the case for the constant potential in subsection \ref{v0}, phase trajectories in $\Psi_+$ are repelled by the segment $(0,1,z)$ (middle figure).}\label{fig02}
\end{center}
\end{figure}

\subsection{The exponential Potential $V=V_0\exp{(-\alpha\phi)}$}\label{expot}

The autonomous system of ordinary differencial equations (\ref{asode}) takes the following form:

\bea
&&x'=-3x+\sqrt\frac{3}{2}\;\alpha\;
 y^2z+\frac{3}{2}x(\gamma(1-x^2-y^2)+2x^2),\nonumber\\
&&y'=-\sqrt\frac{3}{2}\;\alpha\;
 xyz+\frac{3}{2}y(\gamma(1-x^2-y^2)+2x^2),\nonumber\\
&&z'=\frac{3}{2}\;z\left(\frac{z^2-1}{z^2+1}\right)(\gamma(1-x^2-y^2)+2x^2).\label{asode2}\eea Before going any further into the phase space analysis of (\ref{asode2}), let us note that, contrary to the case with the autonomous system of diferential equations (\ref{asode1}), thanks to the term with the factor $\partial_\phi V(\phi)=-\alpha$ in the right-hand-side of the first couple of equations, the equations in (\ref{asode2}) form a coupled system of differential equations. The inmediate consequence is that points in the subsets $(0,\pm 1,z)\in\partial\Psi_\pm$ are not even critical points of the autonomous system (\ref{asode2}).

The critical points of (\ref{asode2}) are summarized in table \ref{tab3}. We have not included the points with $z=-1$, since these coincide with those with $z=1$ under the change of sign of the exponent $\alpha\rightarrow -\alpha$. Recall that the critical points with $z=1$ correspond to the formal limit $r_c\rightarrow\infty$, i. e., these represent standard 4D behaviour. A complete study of the critical points in table \ref{tab3} can be found in the well known reference \cite{wands}. The main results can be summarized as follows:

\begin{itemize}

\item{$\alpha^2<3\gamma$;} The kinetic-dominated solutions (points $P_{2,3}$ in table \ref{tab3}) are unstable nodes (sources). The matter-dominated solution (point $P_1$) is a saddle point. The scalar field dominated solution (point $P_4$) is the late-time attractor.

\item{$3\gamma<\alpha^2<6$;} The kinetic-dominated solutions $P_{2,3}$ are unstable nodes. The scalar field dominated solution $P_4$ is a saddle. The matter-scaling solution (point $P_5$ in table \ref{tab3}) is a stable node/spiral.

\item{$\alpha^2>6$;} The kinetic-dominated solution can be either a source or a saddle. The matter-dominated solution is a saddle point. The matter-scaling solution $P_5$ is a stable spiral.

\end{itemize}

The main difference with the case for the constant potential $V=V_0$ of subsection \ref{v0} is in the absense of the critical submanifolds $M_\pm$ that can be associated with 4D phantom-like behaviour originated from 5D dynamical screening of the potential energy of the scalar field. Although 5D behavior related to dynamical screening does actually arise in the present case (as can be seen from the figure \ref{fig02}, the phase trajectories leave the plane ($x,y,1$) and probe the bulk of the 3D phase space as for the constant potential), the corresponding DGP-Friedmann behaviour dictated by the equation $$H^2+\frac{1}{r_c}\;H=\frac{1}{3}\;V(\phi),$$ is not even a critical point, so that 5D gravitational screening mimetizing 4D phantom behaviour is not a generic solution in this case. Actually, it is evident from the middle figure in Fig. \ref{fig02}, that the trajectories in phase space $\Psi_+$ (the Minkowski cosmological phase) do not even touch the segment $(0,1,z)\in\partial\Psi_+$. In fact, for the chosen initial conditions, these seem to be repelled from that segment.

Another interesting conclusion that can be made after the phase space analysis in this subsection is that, the claim in reference \cite{zhang} that scaling solutions do not exist in a DGP brane filled with dust, in general, is not correct. For $\alpha^2>3\gamma$, matter scaling solutions indeed exist, they are associated with critical points in phase space (point $P_5$ in table \ref{tab3}), and can be, even, attractor solutions.

\section{Self-accelerating solution $H=1/r_c$}\label{selfa}

As already noted from the former analysis, for the self-accelerating phase of the DGP brane model, independent of the initial conditions, trajectories in phase space approach to the self-accelerating solution with phase space coordinate $z=0\;\Rightarrow\;H=1/r_c$. However, the points with $z=0$ are to be removed from the phase space since, otherwise, the coordinates $x$ and $y$ in (\ref{dvars}) are undefined. Therefore, a new choice of phase space coordinates is mandatory to find the stability properties of the self-accelerating solution in phase space. Let us start with the DGP-Friedmann equation for the self-accelerating cosmological phase, written in the following general form

\be H^2-\frac{1}{r_c}\;H=\frac{1}{3}\;\rho,\label{acel1}\ee where $\rho$ is now the energy density of the total matter content trapped on the DGP brane. If new phase space variables are defined:

\be x\equiv\frac{1}{r_c H},\;\;y\equiv\frac{\sqrt\rho}{\sqrt 3 H},\label{acel2}\ee then, the following Friedmann constraint arises:

\be 1=x+y^2\;\Rightarrow\;y=\sqrt{1-x},\label{acel3}\ee so that, to have real valued $y$, the variable $x$ is constrained to the interval $x\in[0,1]$. This, in turn, yields that $y\in[0,1]$ (the case $y\in[-1,0]$ is related with a contracting phase of the cosmic evolution). In consequence the phase space is $\Psi^*_-=\{(x,y):x\in[0,1], y\in[0,1]\}$. Due to the Friedmann constraint (\ref{acel3}), we are left with only one independent autonomous differential equation:

\be x'=3\gamma\; x \left(\frac{1-x}{2-x}\right).\label{acel4}\ee The corresponding critical points are the standard (4D) Friedmann solution $P_1=(0,1)\;\Rightarrow\;H^2=\rho/3$ and the self-accelerating (5D) solution $P_2=(1,0)\;\Rightarrow\;H=1/r_c$. To study the stability we must perturb (\ref{acel4}) in the neighbourhood of the critical points: $x\rightarrow x_i+\epsilon$. Consider linear perturbations, i. e., we neglet terms $\sim {\cal O}(\epsilon^2)$, then

\be \epsilon'=3\gamma\; (x_i+\epsilon)\left(\frac{1-x_i-\epsilon}{2-x_i-\epsilon}\right),\label{acel5}\ee so that, for the Friedmann solution (point $P_1$),

\be \epsilon'=\frac{3\gamma}{2}\epsilon+{\cal O}(\epsilon^2)\;\Rightarrow\;\epsilon(\tau)=\epsilon(0)\;e^{\frac{3\gamma}{2}\;\tau},\label{acel6}\ee while, for the self-accelerating phase:

\be \epsilon'=-3\gamma\epsilon+{\cal O}(\epsilon^2)\;\Rightarrow\;\epsilon(\tau)=\epsilon(0)\;e^{-3\gamma\tau}.\label{acel7}\ee From the analysis of (\ref{acel6}) and (\ref{acel7}) it is evident that the standard Friedmann solution is unstable, while the self-accelerating one is stable against small (linear) perturbations. Otherwise, in the languaje of the dynamical systems analysis, the latter solution is an attractor in phase space. This solution is, precisely, the one corresponding to the point $(x,y,z)=(0,1,0)$ in the former section, that has been removed from the correponding phase space. Therefore, the reason why phase trajectories in $\Psi_-$ of the former section (for both constant and exponential potentials) approach to that point, is that, in adequate phase space coordinates, the self-accelerating solution $H=1/r_c$ is an attractor critical point.

\section{Results and Discussion}

From the analysis in the former sections, the following important results can be summarized:

\begin{itemize}

\item{For the constant potential the first 5 critical points ($P_1$, $P_2^\pm$ and $P_3^\pm$) coincide with those in standard 4D Einstein-Hilbert theory coupled to a self-interacting scalar field with constant self-interaction potential. However, in principle, the phase trajectories are not constrained to the phase plane $(x,y,1)$ as in standard 4D gravity coupled to a self-interacting scalar field, these probe the bulk of the 3D phase space. For solutions in $\Psi_-$, the (standard 4D) stiff-matter solution (critical points $P_2^\pm$ in table \ref{tab1}) are sources (unstable nodes), while the 4D matter-dominated solution (point $P_1$) is a saddle point. Phase trajectories (asymptotically) approach to the point $(0,1,0)$ that has been removed from the phase space to avoid singularities related to the choice of phase space variables $x$, $y$ and $z$. This point, when studied separatelly -- with a different choice of phase space coordinates -- corresponds to the self-accelerating solution $H=1/r_c$, which is stable against small (linear) perturbations, i. e., it is an attractor solution for phase trajectories in $\Psi_-$.}

\item{There are two critical segments $M_\pm=(0,\pm 1,z)$ that belong in the boundary of the Minkowski cosmological phase $\partial\Psi_+$ and exist only for the constant potential case. Critical points in these subsets of the phase space correspond to inflationary solutions and are non-hyperbolic. Only stable and centre submanifolds can be attached to them. A strightforward analysis of the autonomous system of ordinary differential equations (\ref{asode1}), shows that the stability of these critical subspaces depends on the way phase space trajectories approach to them. Phase portraits, however, reveal that points in these segments are seen by the corresponding phase trajectories as sinks or attractor points.}

\item{In general, the dynamical behaviour of the DGP model with exponential self-interaction potential, does not differ from the standard behaviour of 4D Einstein-Hilbert gravity coupled to a self-interacting scalar field (see reference \cite{wands}), but for the fact that trajectories in phase space, inevitably, leave the phase plane ($x,y,1$) and probe the bulk of the 3D phase space due to the 5D nature of the DGP model. In this case, however, there are not critical points that can be associated with (5D) gravitational screening of the scalar field energy density.}

\end{itemize}

Perhaps, the most interesting finding of the present investigation can be associated with the existence of the critical submanifolds $M_\pm=(0,\pm 1,z\in [1,\infty[)$, that are distinctive of the constant potential case only. The critical points in $M_\pm$ are associated with dynamical screening of the cosmological constant $V_0$ and, since these are attractors in $\Psi_+$, the fate of the associated cosmological evolution is quite generic and independent of the initial conditions. Therefore, the existence of $M_\pm$ serves as a nice illustration of dynamical screening of the cosmological constant (a phenomenom of 5D nature) in phase space. The fact that, for the exponential potential case, the semi-infinite segments $(0,\pm,z)\in\partial\Psi_+$ are not even critical subspaces, came as a surprise. The inmediate consequence is that, for the exponential potential, the gravitational screening is not as generic as for the constant potential. 

\begin{figure}[t]
\begin{center}
\includegraphics[width=5.5cm,height=5cm]{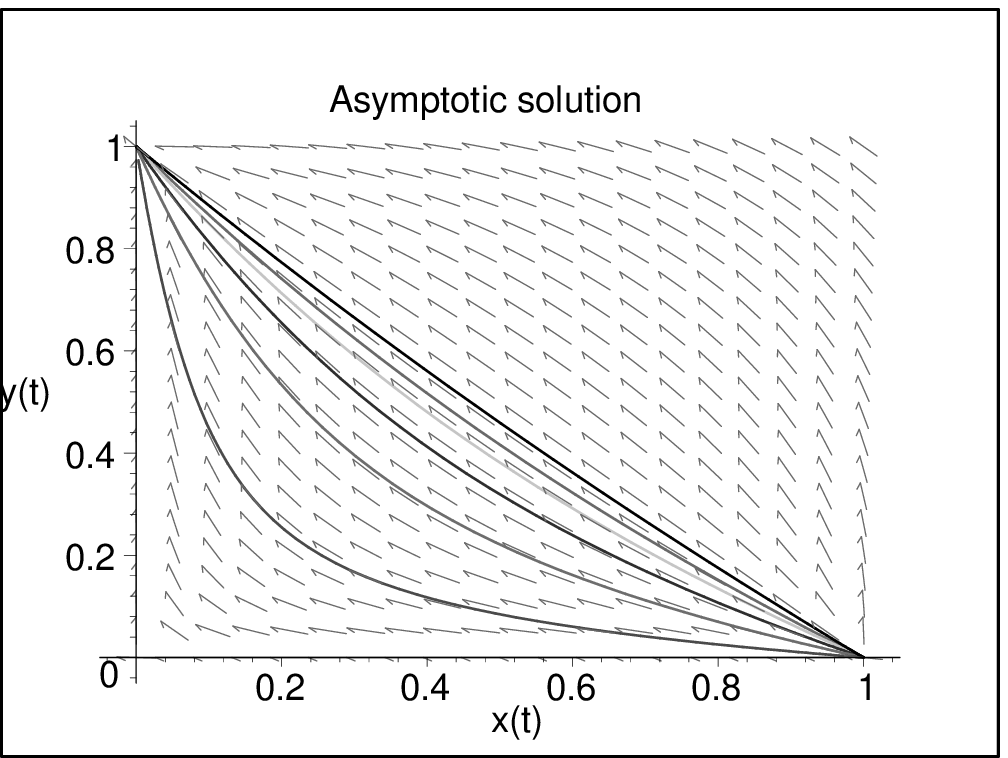}
\includegraphics[width=5.5cm,height=5cm]{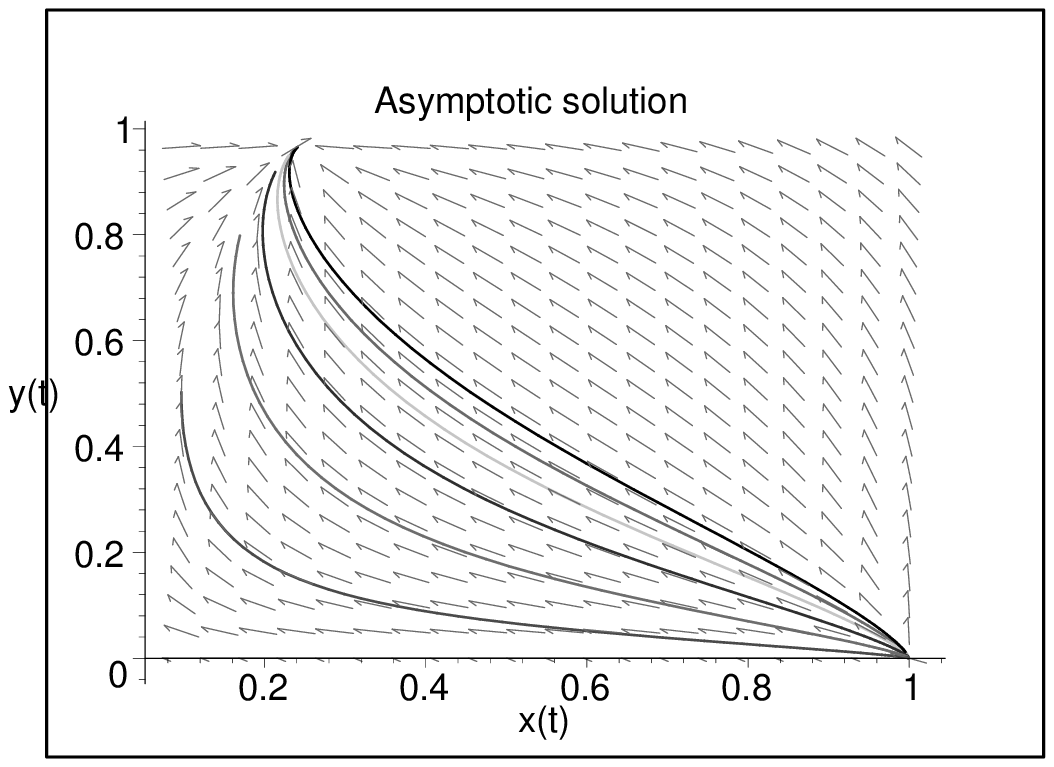}
\vspace{0.3cm}
\bigskip
\caption{Projection of phase trajectories onto a given plane $(x,y,z_0)$ for the Minkowski phase of the DGP model, for the constant potential (upper figure), and for the exponential potential (lower figure), respectively. We chose an arbitrary $z_0=2$. Notice that, while the point $(x,y)=(0,1)$ is an attractor for $V=V_0$ (upper figure), for the case $V=V_0\exp(-\alpha\phi)$ (lower figure), it is not.} \label{fig03}
\end{center}
\end{figure}

The latter conclusion is nicely illustrated in the figure \ref{fig03}, where phase trajectories have been projected onto a given phase plane $(x,y,z_0)$. It is seen that, for the case $V(\phi)=V_0$, trajectories in phase space converge towards the point $(0,1)$ (recall that we fixed $z=z_0$), while for $V(\phi)=V_0\exp(-\alpha\phi)$ these trajectories are repelled by this point. 

The fact that phase trajectories probe the bulk of the 3D phase space $\Psi_\pm$ means that, depending on the initial conditions, the cosmological dynamics of DGP models can be very rich and complicated. A clear illustration of this assertion is associated, precisely, with the occurrence of the attractor (self-accelerating) solution $H=1/r_c$ in $\Psi_-$, and of the critical (also attractor) subspaces $M_\pm\in\partial\Psi_+$, for the constant potential case.

\section{Conclusion}

The dynamics of DGP models can be very rich and interesting. A dynamical systems approach to the subject reveals that trajectories in phase space scape from the plane $(x,y,1)$, that is associated with standard 4D Einstein-Hilbert theory coupled to a self-interacting scalar field. This feature is associated with the 5D nature of the DGP model and originates a phenomenom called as dynamical screening, that mimetizes 4D phantom behaviour. While the latter phenomenom is quite generic and independent of the initial conditions if the self-interaction potential is a constant ($V=V_0$), when $V=V_0\exp(-\alpha\phi)$, the gravitational screening indeed occurs, but it is not a generic phenomenom.

Another interesting result can be associated with the existence of matter-scaling solutions that can have, even, attractor nature. This result is in contradiction with the claim in \cite{zhang} that scaling solutions do not exist in DGP universes with dust. In the case of the exponential potential, the existence of scaling solutions is expected since, the corresponding DGP model is a generalization of the Einstein-Hilbert theory coupled to a (self-interacting) scalar field with $V=V_0\exp(-\alpha\phi)$ (the dynamics of this model has been investigated in reference \cite{wands}), to include 5D effects originated from induced gravity on the brane. Then, since in the formal 4D limit $r_c\rightarrow\infty$ the results of reference \cite{wands} are to be recovered, the critical points found in that reference -- including the scaling solution -- have to be, also, critical points of the corresponding DGP model.

It can be of insterest to investigate the present DGP scenario for arbitrary self-interaction potentials, to show that independence of the gravitational screening of the initial conditions is distinctive only of the constant potential case. This task would entail a different approach than the one undertaken in this paper, so that we leave it for future work.

\section*{ACKNOWLEDGMENTS} 

The authors want to aknowledge very useful comments by Ruth Lazkoz, Genly Leon, Roy Maartens, and Antonio Padilla, on the original version of the present manuscript. This work was partly supported by CONACyT M\'exico, under grants 49865-F, 54576-F, 56159-F, 49924-J, and by grant number I0101/131/07 C-234/07, Instituto Avanzado de Cosmologia (IAC) collaboration. I Q aknowledges also the MES of Cuba for partial support of the research.

\end{document}